\documentclass{article}
\usepackage[utf8]{inputenc}
\usepackage{graphicx}
\usepackage{amsmath}

\title{Geometric Pressure Minimisation of a Theoretical Die for Ceramic Extrusion}
\author{D. J. Leech, S. Lightfoot, T. J{\o}rgensen}
\date{Centre for Fine Print Research, University of the West of England, Bristol, UK}

\begin{document}

\maketitle

\begin{abstract}
    As manufacture processes become more complex, the direct extrusion of material still remains a reliable method of fabrication and processing a variety of materials, but most commonly ceramics. These materials are shaped by a die that can determines the output shape of the material. A generic complex die design was used to explore how varying the geometry of the die can affect the overall and component pressures felt during the extrusion process, based on a Benbow-Bridgwater model. From this we determined some general interactions and relationships, comparing how varying geometries can lead to extrusions of the same shape, and discuss how to balance this against desired effects that the extrusion process may have on the material.
\end{abstract}

\section{Introduction}

Ceramic extrusion is a common processing technique, useful in the production of objects with regular cross-sectional areas and usually the source of building materials such as tiles or bricks \cite{handle_springer_2007}. Due to plastic-like quality of the clay-water mix, ceramics can easily be cold-formed and therefore extrusion has allowed for a simple and achievable route for ceramic formation for decades.

Alongside this, the simplicity of the extrusion process allows for a wide-range of clay formulations, with or without additives, to be utilised. But in general, ceramics are characterised by low thermal and electrical conductivity and high melting temperature. All of these properties are determined entirely by the microstructural mix and can be varied to affect the porosity, mechanical strength and overall shape of the final product.

The extrusion of ceramic materials can be mathematically described using the Benbow model, that relies on both materials properties and the geometry of the extrusion die to describe the pressure drops felt across the die \cite{benbow_amceram_1989}. This was then expanded to include more complex geometries and multi-component dies \cite{benbow_powdertech_1991}. With more current manufacturing capabilities, we can instead begin to imagine more complex and optimal die designs that minimise the overall pressure felt on across the die \cite{holker_intam_2016, oter_optik_2019}. We use a generic complex die design, such as those in Ref. \cite{vitorino_appclaysci_2015}, in order to try and make general statements about design choices that should result in this minimum pressure geometry.

\section{Benbow-Bridgwater Model}

The extrusion of ceramic pastes can be modelled using the Benbow-Bridgwater model and describes the extrusion pressure $P$ as a function of the material properties and the geometrical description of the die. Such a model can be most simply described in a die with a circular cross section with a square entry feed as \cite{benbow_amceram_1989, horrobin_chemeng_1998, wells_matsci_2005}

\begin{equation}
    P = 2(\sigma_{0} + \alpha V^{m}) \ln \left(\frac{D_{0}}{D}\right) + (\tau_{0} + \beta V^{n}) \frac{4L}{D}.
\end{equation}
Here $\alpha$ and $\beta$ are terms that represent flow properties, $n$ and $m$ are generic material dependent terms and $\sigma_{0}$ and $\tau_{0}$ are the bulk yield and wall shear felt by the material. $D_{0}$ and $D$ are diameters of the barrel and die, $L$ is the die-land length. Finally, the pressure variations are driven by the extrudate velocity $V$.

\begin{figure}
\begin{center}
\includegraphics[width=0.88\textwidth]{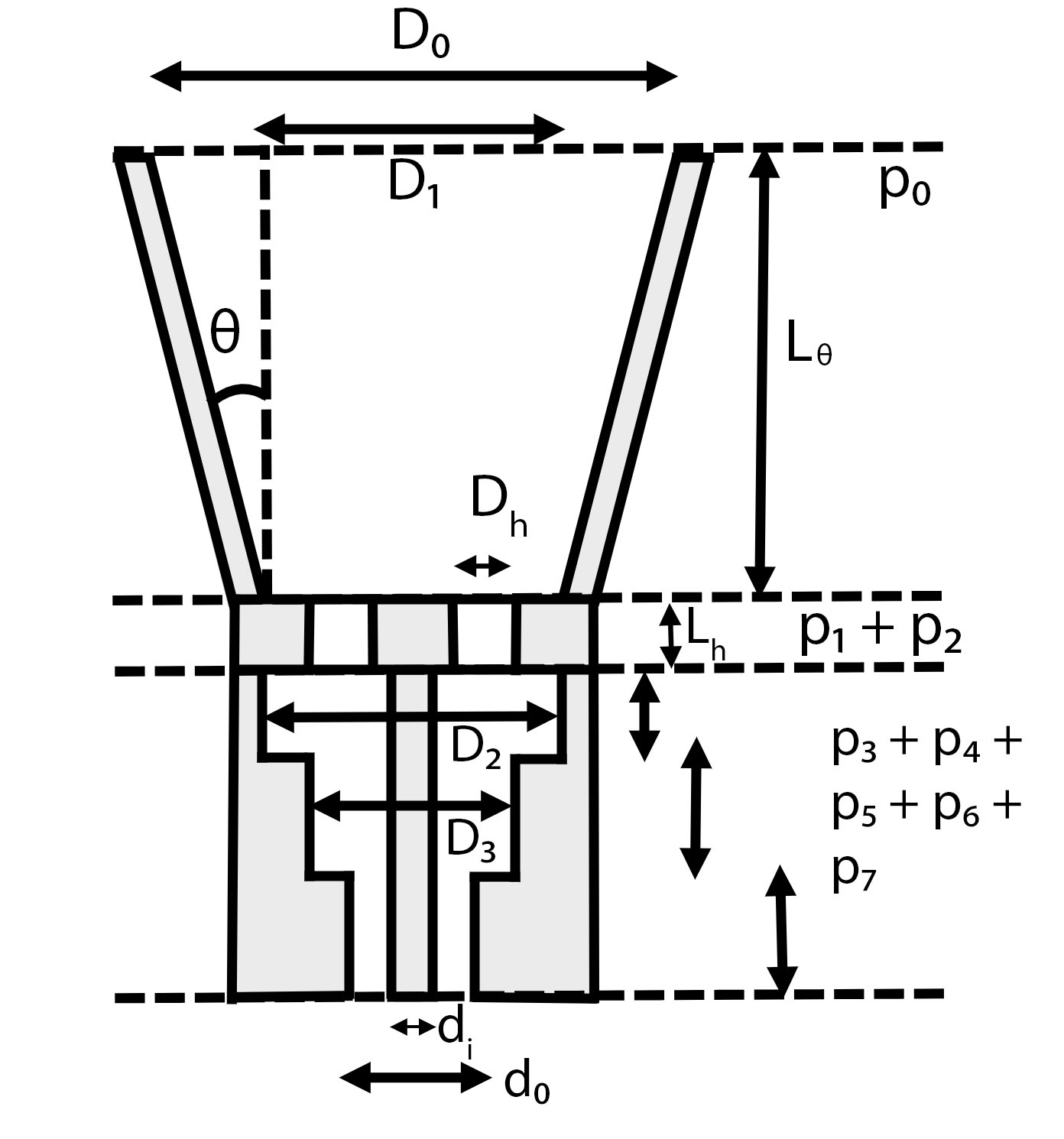} \\
\end{center}
\caption{Representation of the generic complex die design, displaying all the relevant geometrical terms and how the pressure components are split across the length of the die.} \label{DieDesign}
\end{figure}

More complex geometries of die can instead be described by an extension of the above that takes into account the geometry change at each point. The pressure therefore is an additive combination of the pressure drops across each section of the die, with form \cite{vitorino_appclaysci_2015, raupp_euroceram_2007, vitorino_ceramint_2014, ribiero_ceramsoc_2006, guilherme_advceram_2009}

\begin{multline}
\begin{aligned}
    P & = p_{0} + p_{1} + p_{2} + p_{3} + p_{4} + p_{5} + p_{6} + p_{7} \\   
    p_{0} & = \left[ 2 (\sigma_{0} + \alpha V^{m} + \tau_{0} \cot\theta ) \ln \left( \frac{D_{0}}{D_{1}} \right) + \beta V^{n}\cot\theta \right] \\
    p_{1} & = \left[ 2 \left( \sigma_{0} + \alpha \left( \frac{4Q}{\pi D_{h}^{2} N} \right)^{m} \right) \ln\left(\frac{D_{1}}{D_{h}\sqrt{N}} \right) \right] \\
    p_{2} & = \left[ 4 \left( \tau_{0} + \beta \left( \frac{4Q}{\pi D_{h}^{2} N} \right)^{n} \right) \left(\frac{L_{h}}{D_{h}} \right) \right] \\
    p_{3} & = \left[ \left( \tau_{0} + \beta V^{n} \right) \left(\frac{L_{2}M_{2}}{A_{2}} \right) \right] \\
    p_{4} & = \left[ \ln \left(\frac{A_{2}}{A_{3}}\right) (\sigma_{0} + \alpha V^{m} ) \right] \\
    p_{5} & = \left[ (\tau_{0} + \beta V^{n}) \left( \frac{4L_{3}}{D_{3} - d_{i}} \right) \right] \\
    p_{6} & = \left[ \ln \left(\frac{A_{3}}{A_{4}}\right) (\sigma_{0} + \alpha V^{m} ) \right] \\
    p_{7} & = \left[ (\tau_{0} + \beta V^{n}) \left( \frac{4L_{4}}{d_{0} - d_{i}} \right) \right] \\
\end{aligned}
\end{multline}

\begin{figure}
\begin{center}
\includegraphics[width=0.48\textwidth]{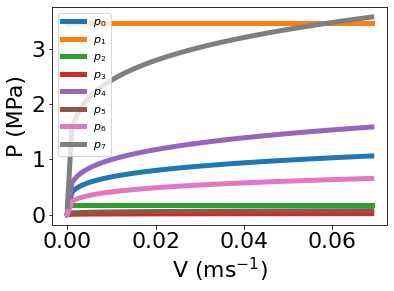}
\includegraphics[width=0.50\textwidth]{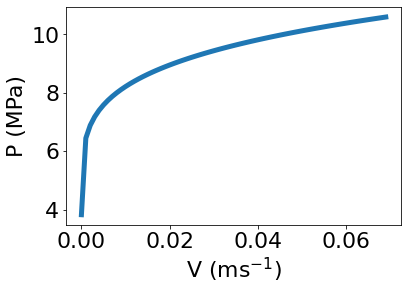} \\
\end{center}
\caption{(a) Individual contributions to the pressure across the various sections of the die with varying extrudate velocity. (b) Total contribution through the summation of all terms. Note that in both plots (and all future plots), the material parameters are provided by the red clay data in \cite{vitorino_appclaysci_2015}.} \label{pInitial}
\end{figure}

$D_{i}$ is the diameter of each portion, $N$ is the number of internal holes with diameter $D_{h}$, $Q$ is the volumetric flow rate, $A_{x}$ and $L_{x}$ are the area and die-land length at a location and $\theta$ is the angle of the die-entry (located solely in the $p_{0}$ term). We use the material example of red clay from Ref. \cite{vitorino_appclaysci_2015}, providing the material dependent parameters ($\tau_{0}$, $\beta$, $n$, $m$, $\alpha$, $\sigma_{0}$). The layout can be seen in Fig. \ref{DieDesign}, with all the geometrical terms above included, and an example of the individual and total contribution to the pressure across the die can be seen in Fig. \ref{pInitial}.

Due to the additive nature of these pressure drops, we can begin this analysis simply by separating the die into smaller sections and noting how differing geometries affect the pressure values. The only fixed terms are the geometry of both the opening ($D_{0}$) and exit (both $d_{0}$ and $d_{i}$) of the die, such that all variations result in the same extruded ceramic shape. It may be that large variations in the opening and exit dimensions may result in a shift of the minimum pressure die geometry, however initial tests suggest that the dependencies described in the following sections hold for at least small variations ($\approx$ 10\%) of these values.

\section{$p_{0}$ - Angular Constrictive Section}

\begin{figure}
\begin{center}
\includegraphics[width=0.80\textwidth]{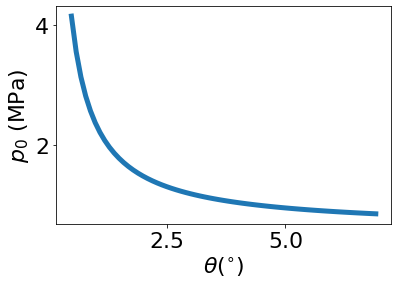} \\
\end{center}
\caption{Value of $p_{0}$ with varying constrictive angle $\theta$. This allows for the length of the die to vary unhindered, locking only the exit diameter $D_{1}$. The general relationship is that as the angle increases, the die-length decreases and so the pressure is reduced. However, for angles above $3^{\circ}$ this change becomes increasingly less important.} \label{AnglePres}
\end{figure}

The region $p_{0}$ belongs to the constrictive section of the die, defined by both an angle of constriction $\theta$ and a length $L_{\theta}$. It is driven purely by:

\begin{equation}
        p_{0} = \left[ 2 (\sigma_{0} + \alpha V^{m} + \tau_{0} \cot\theta ) \ln \left( \frac{D_{0}}{D_{1}} \right) + \beta V^{n}\cot\theta \right] \\
\end{equation}

Varying purely the angle here produces the data seen in Fig. \ref{AnglePres}. It can be seen from this data that an inflexion point lies at $3^{\circ}$ and any angle larger than this is roughly equally preferable when it comes to minimising the pressure felt on the die. This variation can also be thought of as minimising the distance between the barrel and the final extrusion - low angles would provide extremely large distance, with the pressure being asymptotic to $\theta = 0^{\circ}$.

This can be somewhat misleading - the equation does not depend explicitly on the length the die section. As such, variations in the angle have a direct impact on the length and this is not explicitly reflected in the equation. We denote that length $L_{\theta}$ has form

\begin{equation}
    \tan \theta = \frac{D_{0} - D_{1}}{L
    _{\theta}}.
\end{equation}

\begin{figure}
\begin{center}
\includegraphics[width=0.49\textwidth]{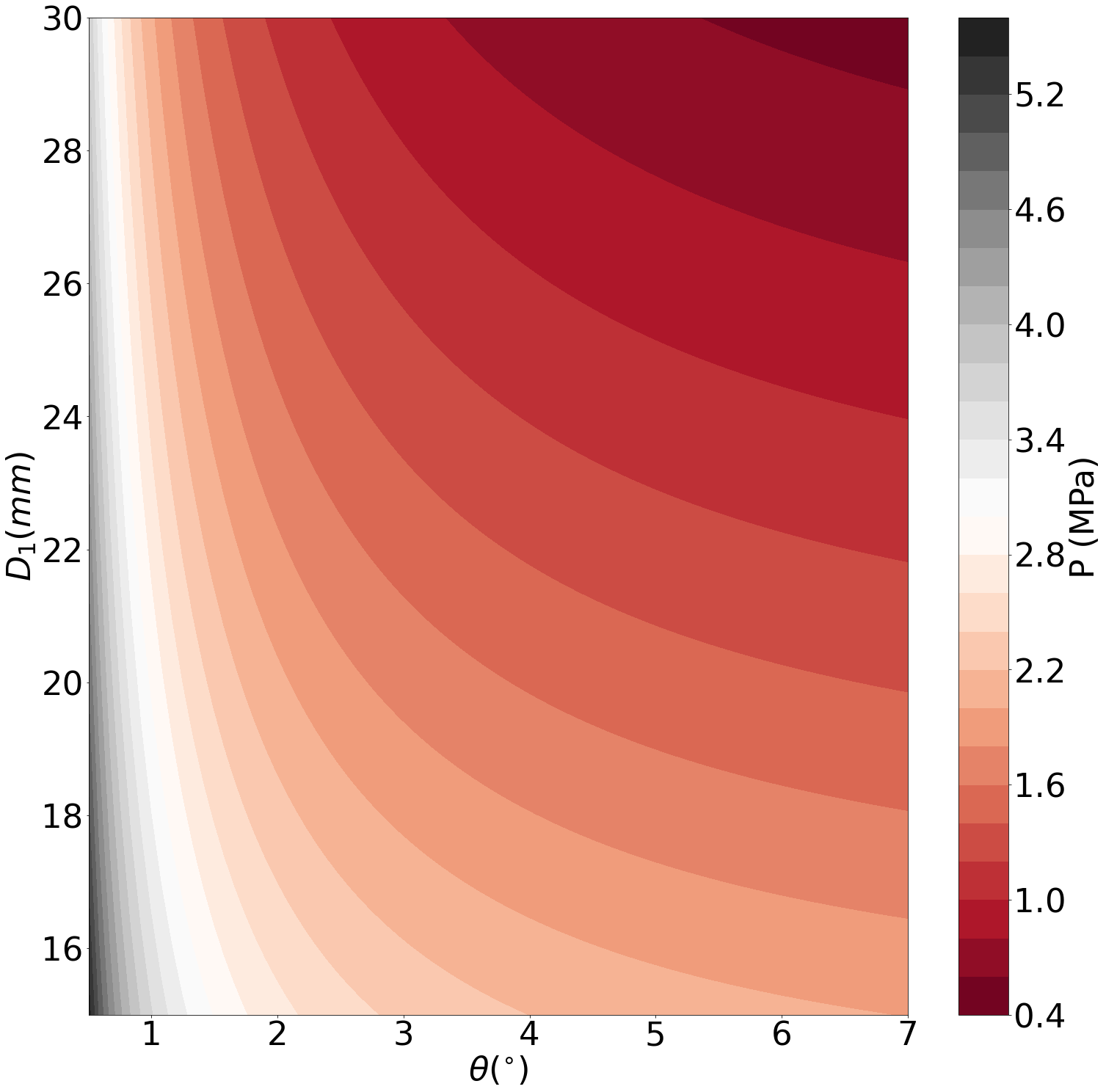}
\includegraphics[width=0.49\textwidth]{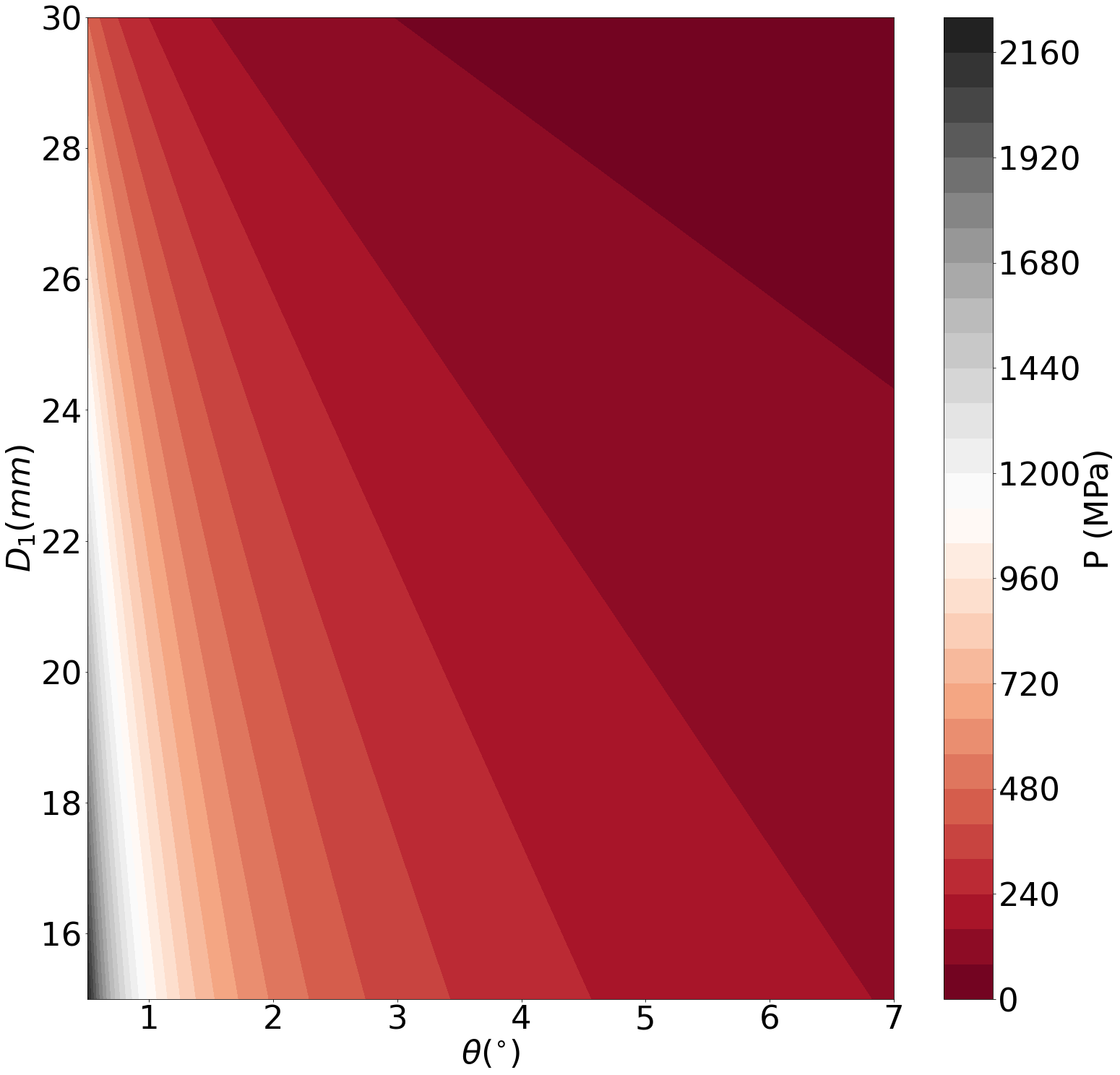} \\
\includegraphics[width=0.80\textwidth]{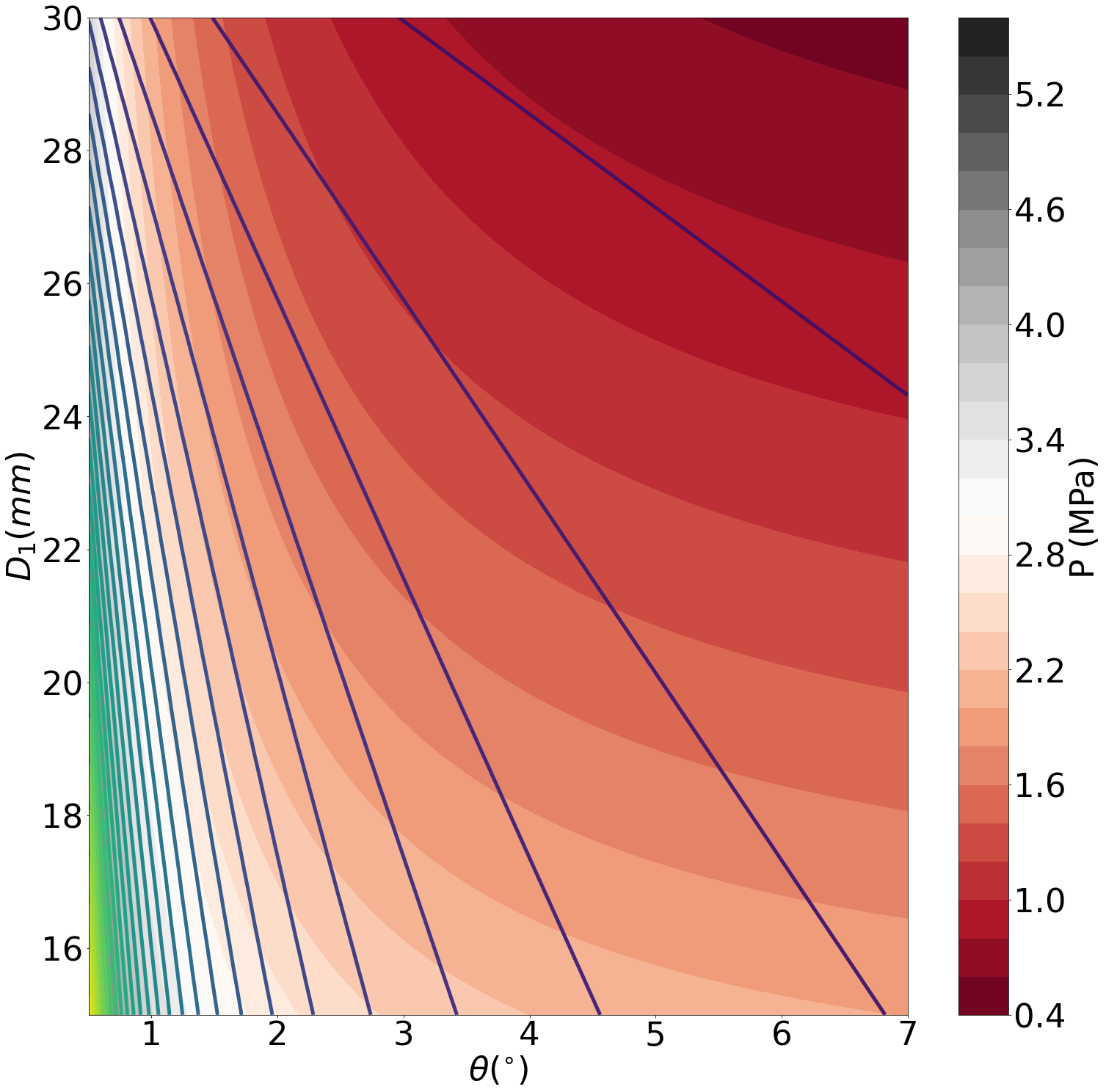}
\end{center}
\caption{(Left) Variation in the pressure with differing die diameter $D_{1}$ and angle between barrel and die $\theta$. The contour colour sections highlight regions of roughly equivalent pressure, such that all combinations in the band would produce the same final result. (Right) Contour instead displaying value of $L_{\theta}$ for combinations of the same parameters. Coloured sections now represents combinations that produce dies of the same length, from barrel to extrusion. (Below) The two plots above overlapped. Now we can choose the minimum pressure geometry, knowing only the desired length of die.} \label{LVary}
\end{figure}

We instead wish to explore how the pressure varies with the exit diameter $D_{1}$ and the angle $\theta$, highlighting regions of equal die-length. Fig. \ref{LVary} shows this data, displaying both the regions of equal pressure felt by the die and regions of equal die length. A combination of these two viewpoints is required in order to minimise the pressure felt, whilst restricting unfeasible cases such as extremely long die lengths or extremely high pressure layouts.

If we overlap these two data sets, we can now produce a figure that allows us the pick the minimum pressure combination of $D_{1}$ and $\theta$, having chosen the length of the desired die. Fig. \ref{LVary} shows this result - indicating regions of geometrical and pressure induced infeasibility. In other words, at large angle $\theta$ and large exit diameter $D_{1}$, the die component length $L_{\theta}$ tends toward zero, and at shallow angles and small exit diameter, the die component length tends towards extremely long die-lengths. It also allows us to state the the minimum pressure setup for a chosen die length, angle or exit diameter. For instance, a die with length $L_{\theta} = 200$ mm, the minimum pressure geometry is roughly in the region of $\theta = 3.5^{\circ}$ and $D_{1} = 26$ mm. There is a region of similar pressure solutions around this point, any of which would provide an ample die geometry to minimise the pressure.

\section{$p_{1}, p_{2}$ - Connector with Internal Holes}

\begin{figure}
\begin{center}
\includegraphics[width=0.49\textwidth]{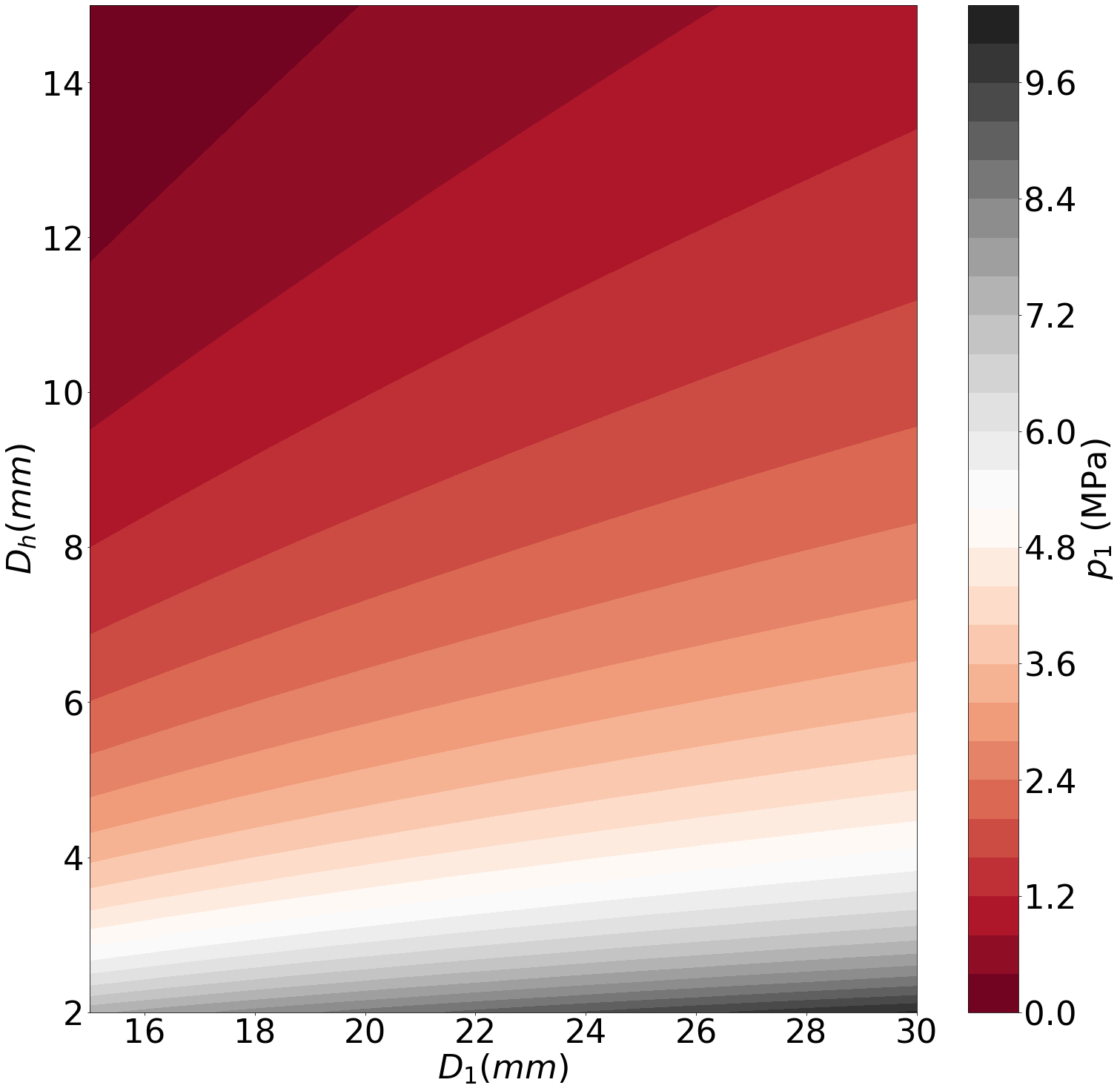}
\includegraphics[width=0.49\textwidth]{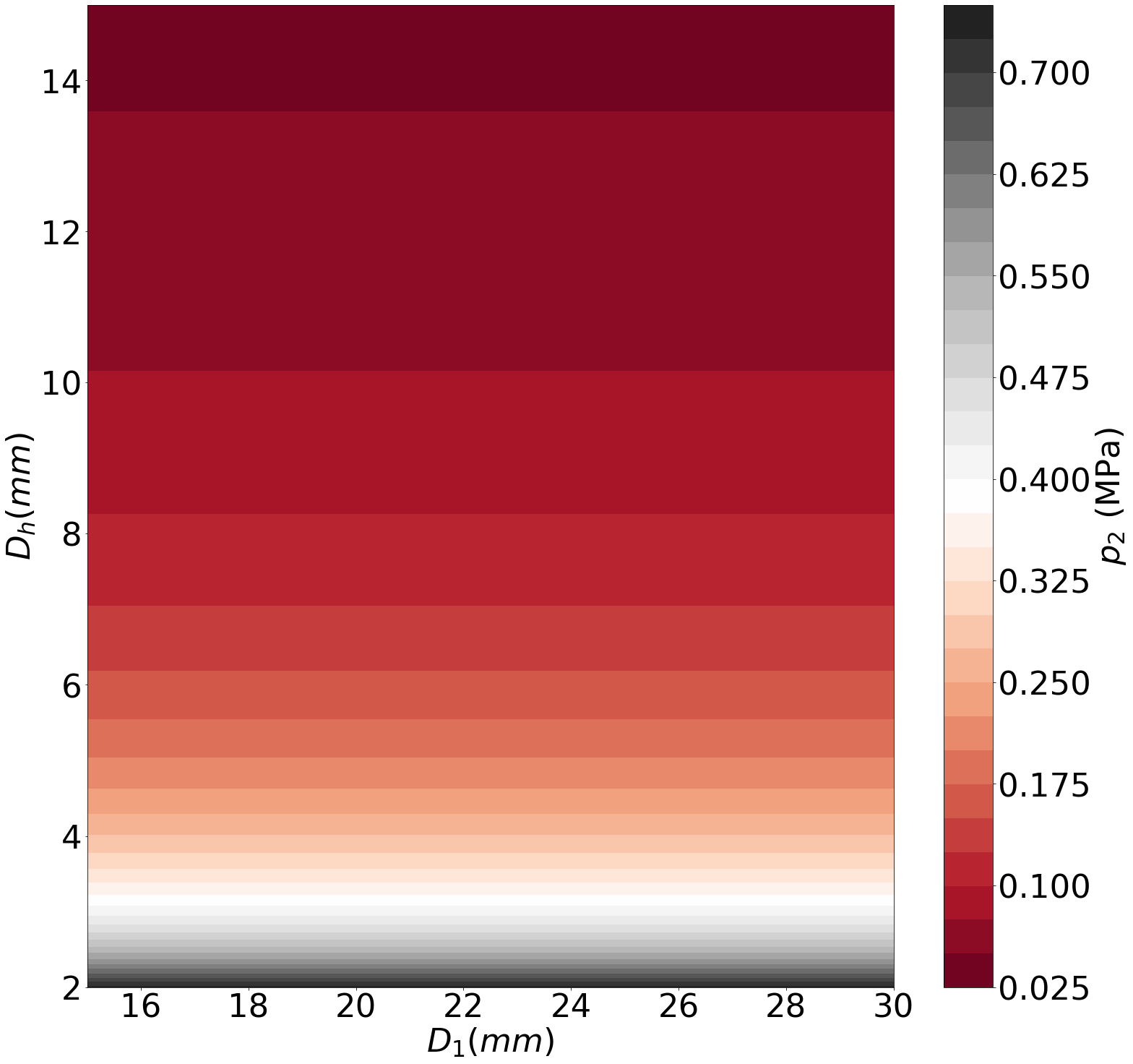}
\end{center}
\caption{Variation in $p_{1}$ (left) and $p_{2}$ (right) in the connector with internal holes portion of the die, with the geometric terms $D_{h}$, hole diameter, and $D_{1}$, connector diameter.} \label{DhD1}
\end{figure}

The following portion is mathematically described by the terms $p_{1}$ and $p_{2}$ - denoting a connector with a series of internal holes that feed to the final portion of the die. This has form
\begin{multline}
\begin{aligned}
    p_{1} + p_{2} & = \left[ 2 \left( \sigma_{0} + \alpha \left( \frac{4Q}{\pi D_{h}^{2} N} \right)^{m} \right) \ln\left(\frac{D_{1}}{D_{h}\sqrt{N}} \right) \right] \\
    & + \left[ 4 \left( \tau_{0} + \beta \left( \frac{4Q}{\pi D_{h}^{2} N} \right)^{n} \right) \left(\frac{L_{h}}{D_{h}} \right) \right]. \\
\end{aligned}
\end{multline}
Here the pressure is directly dependent on the material properties, the die geometry and $Q$, the volumetric flow rate. 

Within this region, using the material parameters for red clay in Ref. \cite{vitorino_appclaysci_2015}, $p_{1}$ is usually much larger than $p_{2}$. Fig. \ref{DhD1} for instance shows immediately this difference and how often $10p_{1} \approx p_{2}$. It also displays how the most obvious geometric terms $D_{h}$, diameter of the internal holes, and $D_{1}$, diameter of the connector between this and the prior section, have a very different effect on the pressure with the former having a profound difference in pressure, while the latter has very little. As such, we can instead state that the more consequential pressure component here is the internal holes in the die. 

\begin{figure}
\begin{center}
\includegraphics[width=0.49\textwidth]{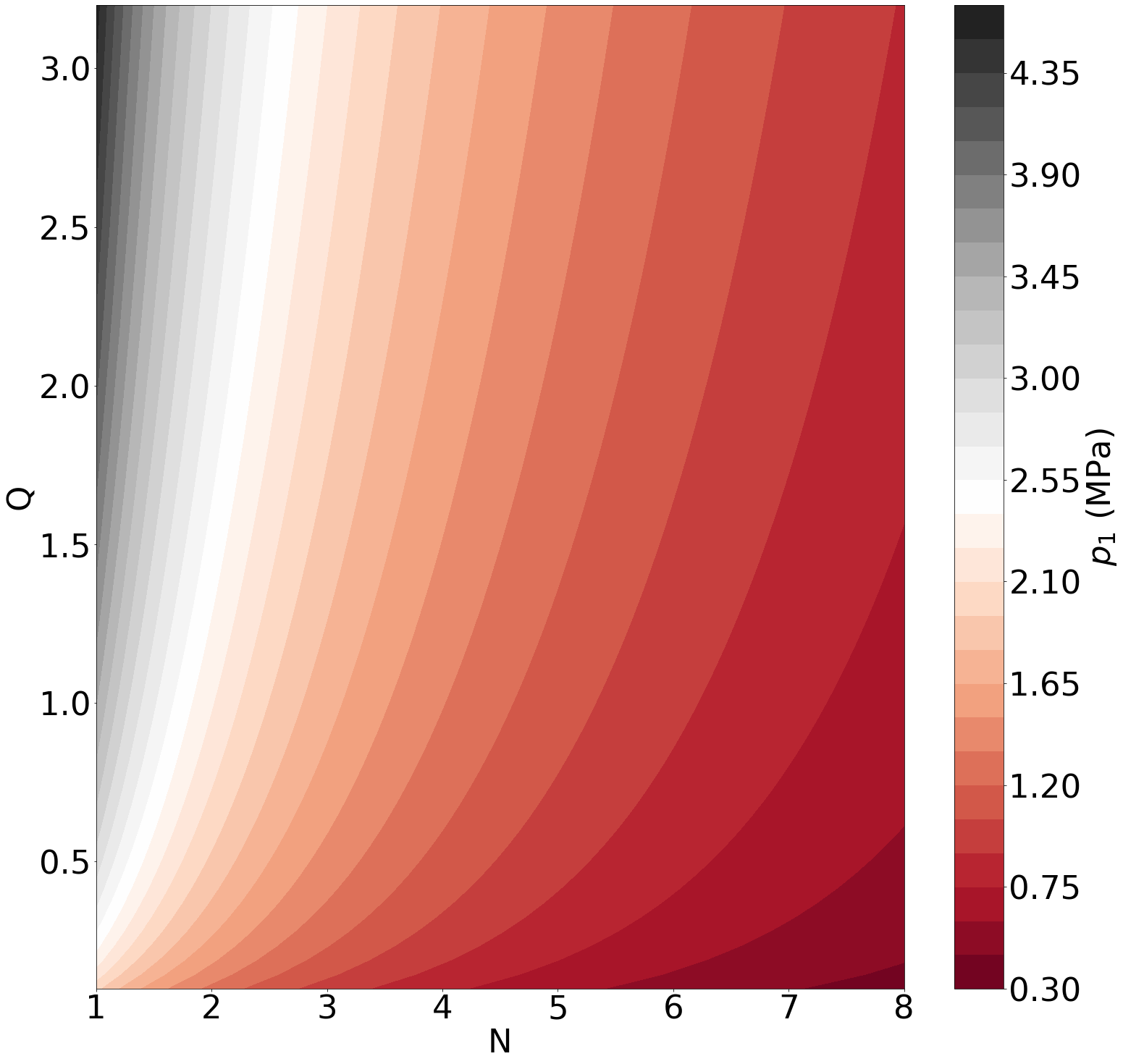}
\includegraphics[width=0.49\textwidth]{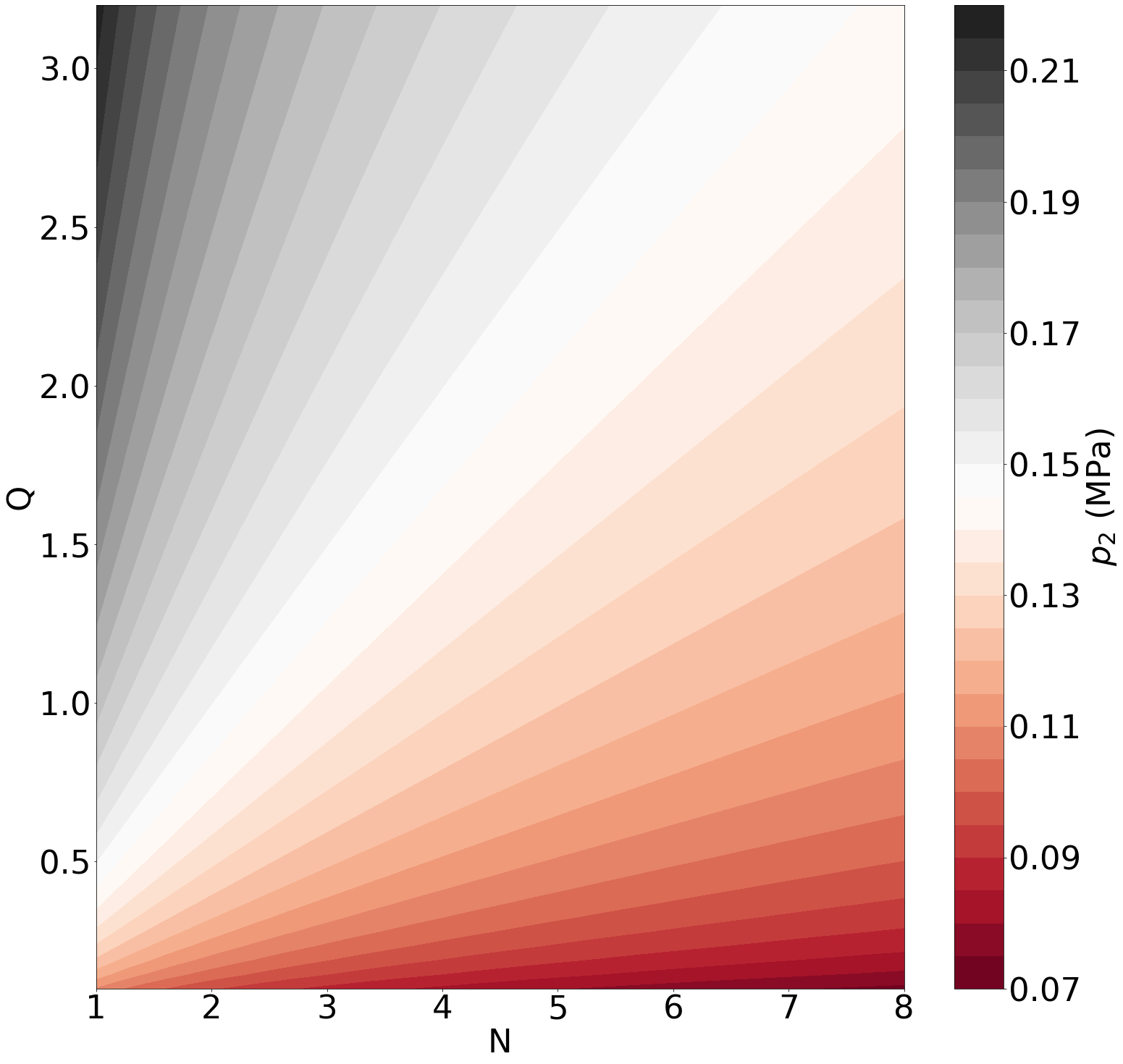}
\end{center}
\caption{Variation in $p_{1}$ (left) and $p_{2}$ (right) in the connector with internal holes portion of the die, with volumetric flow rate $Q$ and number of internal holes $N$.} \label{QN}
\end{figure}

Fig. \ref{QN} shows the pressure variation with volumetric flow rate $Q$ and number of internal holes $N$. As expected, a higher flow rate tends to induce a larger pressure, again with $p_{1} > p_{2}$. Also intuitively, an increased number of holes tends to ensured a reduced pressure. These dependencies are more linear for $p_{2}$ than $p_{1}$, as can be seen from the form of the equations above, however they both highlight how maximising the size of the opening will minimise the pressure. 

This does however highlight one obvious drawback with this method - that it makes no prediction of how the die geometry has an effect on the final extruded material. In certain materials, the extrusion process can have a direct impact on the alignment of the microstructure \cite{peigney_chemphys_2002, oneill_prl_2019, xu_rheo_2020, freitas_appclay_2015, bohm_matsci_1994, kocserha_matscifor_2010} or, in more general cases, cause agglomerate breakdown of ceramic components and agglomeration of defects such as air bubbles \cite{wildman_matsci_1998, mason_matproctech_2009}. The exact purpose that the internal holed structure has may depend on the type of material wanted for extrusion and therefore you easily consider any layout an acceptable pressure loss. However, this also means we can identify regions of similar pressure loss and then choose from those, based on the limitations of our die manufacture.

\section{$p_{3}, p_{4}, p_{5}, p_{6}, p_{7}$ - Stepped Constrictive Section}

\begin{figure}
\begin{center}
\includegraphics[width=0.79\textwidth]{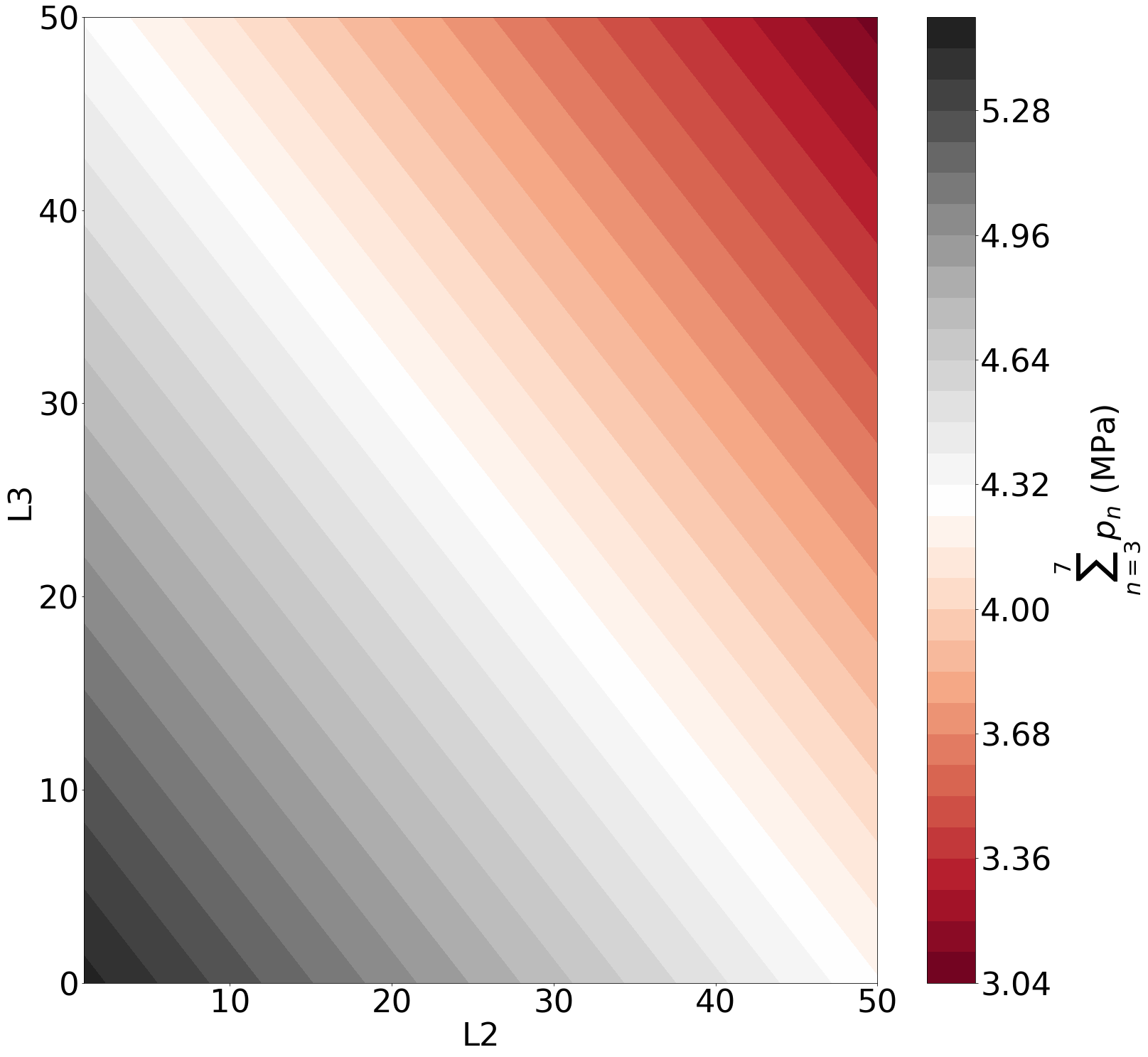}
\end{center}
\caption{$p_{3,4,5,6,7}$ with varying lengths of the stepped section of the die. This is bound by the total length, such that $L_{4} = 120 - L_{3} - L_{2}$.} \label{L2L3}
\end{figure}

The final portion of the die design involves three stepped lengths, characterised by three diameters and three lengths. These have form 

\begin{multline}
\begin{aligned}
    p_{3} + p_{4} & + p_{5} + p_{6} + p_{7} = \\   
    & \left[ \left( \tau_{0} + \beta V^{n} \right) \left(\frac{L_{2}M_{2}}{A_{2}} \right) \right] + \left[ \ln \left(\frac{A_{2}}{A_{3}}\right) (\sigma_{0} + \alpha V^{m} ) \right] \\
    & + \left[ (\tau_{0} + \beta V^{n}) \left( \frac{4L_{3}}{D_{3} - d_{i}} \right) \right] + \left[ \ln \left(\frac{A_{3}}{A_{4}}\right) (\sigma_{0} + \alpha V^{m} ) \right] \\
    & + \left[ (\tau_{0} + \beta V^{n}) \left( \frac{4L_{4}}{d_{0} - d_{i}} \right) \right], \\
\end{aligned}
\end{multline}
where $L_{x}$ and $A_{x}$ are the length and area of their respective sections and $d_{0}$ and $d_{i}$ are the outer and inner diameter of the desired extrusion geometry.

We begin by varying the lengths of the three constituent steps, $L_{2,3,4}$, setting the total length of the section to 120 mm. Fig. \ref{L2L3} shows the pressure relation, constraining the total length through $L_{4} = 120 - L_{3} - L_{2}$. Intuitively, pressure is minimised by increasing the overall average of the diameter of the stepped section, achieved by maximising the lengths of $L_{2}$ and $L_{3}$ over $L_{4}$. Again, this may have adverse effects on the material properties of the extrusion, however this does suggest that similar dimensions of extrusion could be achieved for a much lower pressure. Similar relationships exist for the diameters of the constituent sections, suggesting the pressure can be minimised by maximising the total internal area that still produces the same extrusion shape.

Varying the total length of the die in this section simply increases/decreases the pressure directly, as seen in Ref. \cite{vitorino_appclaysci_2015}, however there is direct evidence that the choice of length here can directly affect the material properties through agglomeration breakdown \cite{bohm_matsci_1994}.

\section{General Results}

Using a generic complex die design, we have explored how varying the geometry of the die can affect the pressure in multiple ways, in differing types of die component. Some broad facts emerge, including:

\begin{itemize}
    \item For constrictive angled die sections, the optimum angle varies depending on the length and opening/exit diameter. However, angles in the region of 3 - 5$^{\circ}$ can in general be a safe choice.
    \item For connectors with internal holes, intuitively the larger the area of entry and the number of entry holes, the lower the pressure. However, this must be matched carefully against the desired effect on the material being fed through connector.
    \item For long, stepped die-land sections, the pressure is minimised by keeping the ceramic in largest sections for as long as possible and by minimising the total length of this component. Again, this is very intuitive but must be matched against the need to affect the material being extruded in this way. Whether this section could be replaced with another angled component and still produce the same material properties is currently unknown.
\end{itemize}

Using this knowledge, we can envisage how custom die-design could result in more bespoke and complex effects on the extruded material, all whilst minimising the overall pressure and wear on such a custom die. In addition, as direct extrusion becomes a method of fabrication for materials beyond ceramics \cite{holker_intam_2016, mohammed_rheo_2017} and more complex die designs become more feasible and accessible, we can also begin to expect material specific die designs.

\section{Conclusion}

A generic complex die was used to probe how varying geometries can affect the overall pressure felt during the extrusion process. From this we determined some general interactions and relationships, primarily how to achieve similar shape outputs in a shorter die. However these do not take into account the effect that the extrusion process has the material itself. This will vary depending on the desired effect and the material chosen, however this does point toward a counterbalancing term that will need to be taken into account for a complete, material specific die design.

\end{document}